\documentclass[prb,twocolumn,showpacs,preprintnumbers,amsmath,amssymb,superscriptaddress,10pt]{revtex4-1}

\usepackage{graphicx}
\usepackage{dcolumn}
\usepackage{bm}

\begin{document}


%
%
%
%
%
%
%
\title{
Field-angle Resolved Flux-flow Resistivity as a Phase-sensitive Probe of Unconventional Cooper Pairing
}

\author{Yoichi Higashi}
\affiliation{Department of Mathematical Sciences, Osaka Prefecture University, 1-1 Gakuen-cho, Naka-ku, Sakai 599-8531, Japan}

\author{Yuki Nagai}
\affiliation{CCSE, Japan Atomic Energy Agency, 5-1-5 Kashiwanoha, Kashiwa, Chiba 277-8587, Japan}

\author{Masahiko Machida}
\affiliation{CCSE, Japan Atomic Energy Agency, 5-1-5 Kashiwanoha, Kashiwa, Chiba 277-8587, Japan}

\author{Nobuhiko Hayashi}
\affiliation{Nanoscience and Nanotechnology Research Center (N2RC), Osaka Prefecture University, 1-2 Gakuen-cho, Naka-ku, Sakai 599-8570, Japan}

\date{\today}

\begin{abstract}
We theoretically investigate the applied magnetic field-angle dependence
of the flux-flow resistivity $\rho_{\rm f}(\alpha_{\rm M})$ for a uniaxially anisotropic Fermi surface.
$\rho_{\rm f}$ is related to the quasiparticle scattering rate inside a vortex core,
which reflects the sign change in the superconducting pair potential. 
We find that $\rho_{\rm f}(\alpha_{\rm M})$ is sensitive
to the sign change in the pair potential
and has its maximum when the magnetic field is parallel to the gap-node direction.
We propose the measurement of the field-angle dependent oscillation of $\rho_{\rm f}(\alpha_{\rm M})$
as a phase-sensitive field-angle resolved experiment.
\end{abstract}
\maketitle
\section{introduction}
It is of great importance to elucidate the symmetry of a superconducting pair potential is of great importance
when studying the Cooper pairing mechanism in unconventional superconductors (SCs).

The pair potential is composed of
spin and orbital wave functions.
The orbital wave function is characterized by its amplitude and phase (sign of the wave function).

In the past decade,
experimental techniques
for the field-angle resolved
specific heat and thermal conductivity measurements
have developed to identify the Cooper pairing symmetry
in various superconducting systems.\cite{Sakakibara}
These angle-resolved mesurements are powerful techniques
that can detect the anisotropy of the pair potential amplitude.
The theory proposed by Vorontsov and Vekhter has successfully explained these experiments for CeCoIn$_5$ assuming $d$-wave Cooper pairing.\cite{vorontsov2007}
However,
these field-angle resolved experiments cannot probe the sign change in the pair potential. 
That is,
they are not phase-sensitive probes.
In addition to detecting the anisotropy of the pair potential,
it is crucial to probe the phase of the Cooper pair
in order to discriminate unconventional SCs,
including iron-based SCs,
from conventional ones.

Until now, only a few phase-sensitive probes have been developed and succeeded,
e.g., the half-flux quantum observation in the tricrystal geometry 
by a scanning SQUID (superconducting quantum interference device) microscope,\cite{Tsuei1994} 
and detecting the quasiparticle interference pattern 
by scanning tunneling spectroscopy (STS).\cite{Hanaguri2010}
Another phase-sensitive probe is measuring bound states at an interface
by point-contact spectroscopy or STS experiments.
If both sides of a SC/SC junction are of the same pair potential amplitude but with opposite signs,
the quasiparticle (QP) path through the interface acquires a $\pi$ phase shift and generating bound states around the interface.
This situation is similar to a vortex line in superconductors.
However, there is the difficulty of fabricating a junction in terms of nono-structured processing techniques.
Actually, the phase-sensitive test using SC/SC junctions succeeds only for cuprate superconductors.\cite{Tsuei2000}
In addition to these existing experiments, a new phase-sensitive test is highly desired.

In this paper,
we propose a new experiment that can detect the phase (sign-change) of the pair potential free from fabricating a SC/SC junction. 
This is the great advantage of the phase-sensitive test proposed in the present work. 
We theoretically study the in-plane field-angle dependence
of the flux-flow resistivity $\rho_{\rm f}(\alpha_{\rm M})$
for typical gap functions and Fermi surface (FS). 
From our numerical calculations,
we show that the phase-sensitive QP scattering inside a vortex core
leads to different behavior of $\rho_{\rm f}(\alpha_{\rm M})$
between conventional and unconventional Cooper pairing. 
In addition,
we show that
$\rho_{\rm f}(\alpha_{\rm M})$ has its maximum
when the applied magnetic field $\bm{}$ is parallel to the gap-node directions.
Our results show that
the field-angle dependence of the flux-flow resistivity
can detect both the sign change of the pair potential and the direction of the gap nodes.


\section{flux-flow resistivity and quasiparticle scattering rate}
The flux-flow resistivity $\rho_{\rm f}$ is
dominated by the quasiparticle within a vortex core.
We assume the system belongs to the moderately clean regime,
in which there are two important contributions to $\rho_{\rm f}$.
One is the QP scattering rate $\varGamma$ inside a vortex core,
and the other is the momentum-dependent quantum level spacing
of the vortex bound states $\omega_0(\bm{k}_{\rm F})$.\cite{kopnin1997,makhlin1997}
Here,
the QP scattering is due to non-magnetic impurities
randomly distributed in the system.

\begin{figure}[tb]
  \begin{center}
    \begin{tabular}{p{80mm}}
      \resizebox{80mm}{!}{\includegraphics{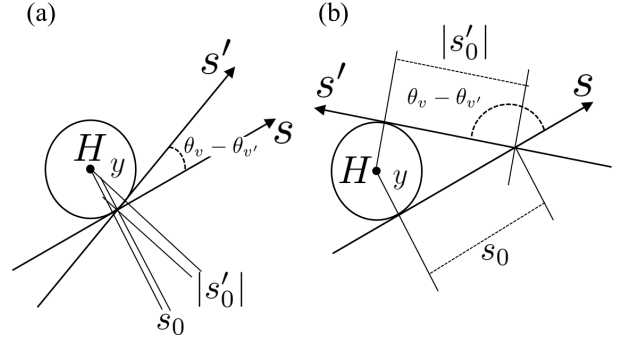}}\\
    \end{tabular}
\caption{
\label{fig:fig1}
The schematic figures of (a) the forward scattering and (b) the backward scattering in the vicinity of a vortex.
$s$ and $s^\prime$ indicate the QP trajectory before and after scattering, respectively.
} 
  \end{center}
\end{figure}

We attribute the flux-flow resistivity $\rho_{\rm f}$
to the energy dissipation of the vortex bound states
due to the impurity scattering inside a vortex core.\cite{Kato}
$\rho_{\rm f}$ is characterized by the two quantities mentioned above,\cite{kopnin1997,makhlin1997}
\begin{equation}
\rho_{\rm f}(T)
\propto
\cfrac{\varGamma_{\rm n}}{\varDelta_0}
\left[
\cfrac{1}{\nu_0}
\int \cfrac{dS_{\rm F}}{|\bm{v}_{\rm F}(\bm{k}_{\rm F})|}
\cfrac{\omega_0(\bm{k}_{\rm F})}{\varDelta_0}\cfrac{\varGamma_{\rm n}}{\varGamma(\varepsilon=k_{\rm B}T,\bm{k}_{\rm F})}
\right]^{-1},
\label{flux-flow_resistivity}
\end{equation}
where $\varGamma_{\rm n}$ is the impurity scattering rate in the normal state and
$\varDelta_0$ is the bulk amplitude of the pair potential.
We assume that the temperature $T$ dependence of $\rho_{\rm f}$ comes
predominantly from $\varGamma$
with the QP energy $\varepsilon=k_{\rm B}T$.
Here, we have made a rough estimate.
Actually,
the QPs distributed with the energy width $\Delta \varepsilon \sim k_{\rm B} T$
contribute to $\varGamma$.
The total density of states on a FS is
$\nu_0=\int dS_{\rm F}/|\bm{v}_{\rm F}(\bm{k}_{\rm F})|$,
with $dS_{\rm F}=|\bm{k}_{\rm F}(\phi_k,\theta_k)|^2 \sin \theta_k d\phi_k d\theta_k$
being an area element on the FS,
the Fermi velocity
$\bm{v}_{\rm F}(\bm{k}_{\rm F})=\bm{\nabla}_{\bm{k}}\epsilon(\bm{k})|_{\bm{k}=\bm{k}_{\rm F}}$,
and the Fermi wave vector
$\bm{k}_{\rm F}
=
|\bm{k}_{\rm F}(\phi_k,\theta_k)|
(\bar{\bm{a}}\cos \phi_k \sin \theta_k
+\bar{\bm{b}}\sin \phi_k \sin \theta_k
+\bar{\bm{c}}\cos \theta_k)$.
$\epsilon(\bm{k})$ is the energy dispersion of electrons.
$\phi_k$ $(\theta_k)$ is the azimuthal (polar) angle on the FS.
$\bar{\bm{a}}$, $\bar{\bm{b}}$, and $\bar{\bm{c}}$ denote
orthogonal unit vectors spanning crystal coordinates.
We use the unit system in which $\hbar=1$.
The momentum-dependent inter-level spacing of
the vortex bound states $\omega_0(\bm{k}_{\rm F})$
is obtained analytically as \cite{higashijpcs,Caroli}
$
\omega_0(\bm{k}_{\rm F})
=
2 |d(\bm{k}_{\rm F})|^2\varDelta^2_0
/(|\bm{k}_{\rm F \perp}| |\bm{v}_{\rm F \perp}(\bm{k}_{\rm F})|)
$
using the quasiclassical Green's function method
and the Kramer-Pesch approximation.\cite{Nagai2008,Nagai2010}
$d(\bm{k}_{\rm F})$ indicates the anisotropy of pair potential and
the vector with $\perp$ denotes the vector component projected onto the plane perpendicular to $\bm{H}$.
We treat the non-magnetic impurity scattering by means of the Born approximation.\cite{Kato,Buchholtz}
The quasiparticle scattering rate
for the QPs with the energy $\varepsilon$
inside a vortex core
is obtained as \cite{Nagai2010,higashi2011}
\begin{align}
\frac{\varGamma(\varepsilon)}{\varGamma_{\rm n}}
&=
\Bigg\langle
\Bigg\langle
\frac{\varGamma(\bm{k}_{\rm F},\bm{k}^\prime_{\rm F},\varepsilon)}{\varGamma_{\rm n}}
\Bigg\rangle_{\rm FS^\prime}
\Bigg\rangle_{\rm FS},
\label{qp_scattering_rate}
\\
\frac{\varGamma(\bm{k}_{\rm F},\bm{k}^\prime_{\rm F},\varepsilon)}{\varGamma_{\rm n}}
&= \frac{\pi}{2}  
C(\bm{k}_{\rm F},\bm{k}^\prime_{\rm F})
D(\bm{k}_{\rm F},\bm{k}^\prime_{\rm F})
F(\varepsilon,\bm{k}_{\rm F},\bm{k}^\prime_{\rm F}),
\label{k-dep-qp_scattering_rate}
\\
C(\bm{k}_{\rm F},\bm{k}^\prime_{\rm F})
&=
1-\mathop{\mathrm{sgn}}\nolimits[d(\bm{k}_{\rm F})d(\bm{k}^\prime_{\rm F})] \cos\Theta,
\label{coherence-factor}
\\
D(\bm{k}_{\rm F},\bm{k}^\prime_{\rm F})
&=
\frac{1}{\vert\sin\Theta\vert},
\\
F(\varepsilon,\bm{k}_{\rm F},\bm{k}^\prime_{\rm F})
&=
\frac{\vert \bm{v}_{\rm F\perp}(\bm{k}^\prime_{\rm F}) \vert}{\vert \bm{v}_{\rm F\perp}(\bm{k}_{\rm F}) \vert}
\frac{\vert d(\bm{k}_{\rm F}) \vert}{\vert d(\bm{k}^\prime_{\rm F})\vert}
e^{-u(s_0,\bm{k}_{\rm F})}
e^{-u(s^\prime_0,\bm{k}^\prime_{\rm F})}.
\label{F}
\end{align}
Here,
$\langle \cdots \rangle_{\rm FS} \equiv (1/\nu_0)\int dS_{\rm F} \cdots /|\bm{v}_{\rm F}(\bm{k}_{\rm F})|$,
$\Theta(\bm{k}_{\rm F}, \bm{k}^\prime_{\rm F})
\equiv
\theta_v(\bm{k}_{\rm F})-\theta_{v^\prime}(\bm{k}^\prime_{\rm F})$
for the scattering angle  [see Fig.~\ref{fig:fig1}].
$\varGamma$ has the decay factor $\exp[-u(s_0,\bm{k}_{\rm F})]$
with
$u(s_0,\bm{k}_{\rm F})
=
(2\vert d(\bm{k}_{\rm F}) \vert/
\vert \bm{v}_{\rm F \perp}(\bm{k}_{\rm F}) \vert)
\int_0^{\vert s_0 \vert}ds^\prime
\tilde{\varDelta}(s^\prime).
$
We adopt 
$\tilde{\varDelta}(s^\prime)=\varDelta_0 \tanh(s^\prime/\xi_0)$
as the spatial variation of the pair potential.
The coherence length is defined by $\xi_0=v_{\rm F \perp}/(\pi \varDelta_0)$ with
$v_{\rm F \perp} \equiv \langle |\bm{v}_{\rm F \perp}(\bm{k}_{\rm F})| \rangle_{\rm FS}$.
Here,
we define the field-angular dependent effective coherence length for the later discussions as
\begin{equation}
\xi_{\rm eff}(\bm{k}_{\rm F})=|\bm{v}_{\rm F \perp}(\bm{k}_{\rm F})|/[\varDelta_0 |d(\bm{k}_{\rm F})|].
\label{xi_eff}
\end{equation}
This length scale $\xi_{\rm eff}(\bm{k}_{\rm F})$ characterizes the size of the bound states of the QP with the momentum $\bm{k}_{\rm F}$.
Figure~\ref{fig:fig1} shows the QP trajectories
on the plane perpendicular to the magnetic field $\bm{H}$.
The quantities with a prime are those after scattering.
$s_0$ and $|s^\prime_0|$ denote the length
between the point that is the nearest from the vortex center on the QP trajectory
and the scattering point.\cite{Nagai2010,higashi2011}

\section{system}
In this study,
we consider the case in which
$\bm{H}$ is applied parallel to the $a$-$b$ plane and rotated.
The field angle measured from $a$ axis is $\alpha_{\rm M}$.
Here, $a$, $b$, and $c$ are crystal axes.
When calculating the dependence of $\rho_{\rm f}$
on the magnetic field angle $\alpha_{\rm M}$,
we need a coordinate system fixed to $\bm{H}$
which is spanned by
$\bar{\bm a}_{\rm M}$,
$\bar{\bm b}_{\rm M}$,
and $\bar{\bm c}_{\rm M}$
(vortex coordinate system).
Here, 
these axes are orthogonal unit vectors
with $\bar{\bm c}_{\rm M}$
set parallel to 
$\bm{H}$ ($\bar{\bm c}_{\rm M} \parallel \bm{H}$).
$\bm{v}_{{\rm F}\perp}$, $\theta_v$, and those with a prime
are defined in the vortex coordinates.
However,
$\bm{k}_{\rm F}$ and $\bm{k}^\prime_{\rm F}$
are identified by $(\phi_k,\theta_k)$ on a FS
in the crystal coordinates
spanned by
$\bar{\bm{a}}$, $\bar{\bm{b}}$ and $\bar{\bm{c}}$,
which characterize the crystal axes.
In order to calculate the field-angle $\alpha_{\rm M}$ dependence of $\rho_{\rm f}$,
we need to derive the relation between
$\bm{v}_{\rm F \perp}$, $\theta_v$,
and
$\bm{v}_{\rm F}$, $\theta_k$ \cite{higashi2011}.
Then,
the component of $\bm{v}_{{\rm F}}(\bm{k}_{\rm F})$
projected onto the plane perpendicular to $\bm{H}$
is finally obtained as
\begin{eqnarray}
|\bm{v}_{{\rm F}\bot} (\phi_k,\theta_k)|
&=&
|\bm{v}_{\rm F}(\phi_k,\theta_k)|
\Omega(\phi_k,\theta_k),
\label{eq:5}
\\
\Omega(\phi_k,\theta_k)
&=&
\sqrt{
\cos^2\theta_k + \sin^2 (\phi_k-\alpha_{\rm M}) \sin^2\theta_k
},
\label{eq:6}
\\
\cos\theta_v (\phi_k,\theta_k)
&=&
\frac{ -|\bm{v}_{\rm F}(\phi_k,\theta_k)| }
{ |\bm{v}_{{\rm F}\bot} (\phi_k,\theta_k)| }
\cos\theta_k,
\label{eq:7}
\\
\sin\theta_v(\phi_k,\theta_k)
&=&
\frac{ |\bm{v}_{\rm F}(\phi_k,\theta_k)| }
{ |\bm{v}_{{\rm F}\bot} (\phi_k,\theta_k)| }
\sin(\phi_k-\alpha_{\rm M}) \sin\theta_k.
\label{eq:8}
\end{eqnarray}
Thus, the relation between the vortex coordinate and the crystal coordinate
is derived.
Here we give the expression of the projected Fermi velocity for an arbitrary anisotropic FS (see Ref.~\onlinecite{higashi-iss2011} for the expression for a uniaxially anisotropic FS).
We have now reached the position
where we can perform the numerical integration on FS
and calculate the field-angle dependence of Eq.\ (\ref{flux-flow_resistivity}).

We consider the following two types of the simple pair potential model.
One is a line-node $s$-wave pair: $d(\bm{k}_{\rm F})=\vert \cos 2\phi_k \vert \sin^2 \theta_k$. 
The other is a $d_{x^2-y^2}$-wave one: $d(\bm{k}_{\rm F})=\cos 2\phi_k \sin^2 \theta_k$. 
Each one has gap nodes from the north pole of the FS to the south one
in the $\phi_k=(1+2n)\pi/4~[{\rm rad}]$ directions (gap-node directions) with $n=0,1,2,3$.
$\phi_k=n\pi/2~[{\rm rad}]$ directions correspond to anti-node directions.
In the momentum space,
these two pair potentials have the same anisotropy
in their amplitude $\vert d(\bm{k}_{\rm F}) \vert$.
However, only the $d$-wave pair has the sign change
and the $s$-wave pair does not. 

Our calculations are performed for
a uniaxially anisotropic FS
with the mass anisotropy parameter $\gamma=\sqrt{m_c/m_{ab}}$ \cite{higashi-iss2011}. 
$m_c$ and $m_{ab}$ are masses characterizing charge transport
along the $c$-axis and within the $a-b$ plane, respectively.

\begin{figure}[tb]
  \begin{center}
    \begin{tabular}{p{70mm}p{70mm}}
      \resizebox{70mm}{!}{\includegraphics{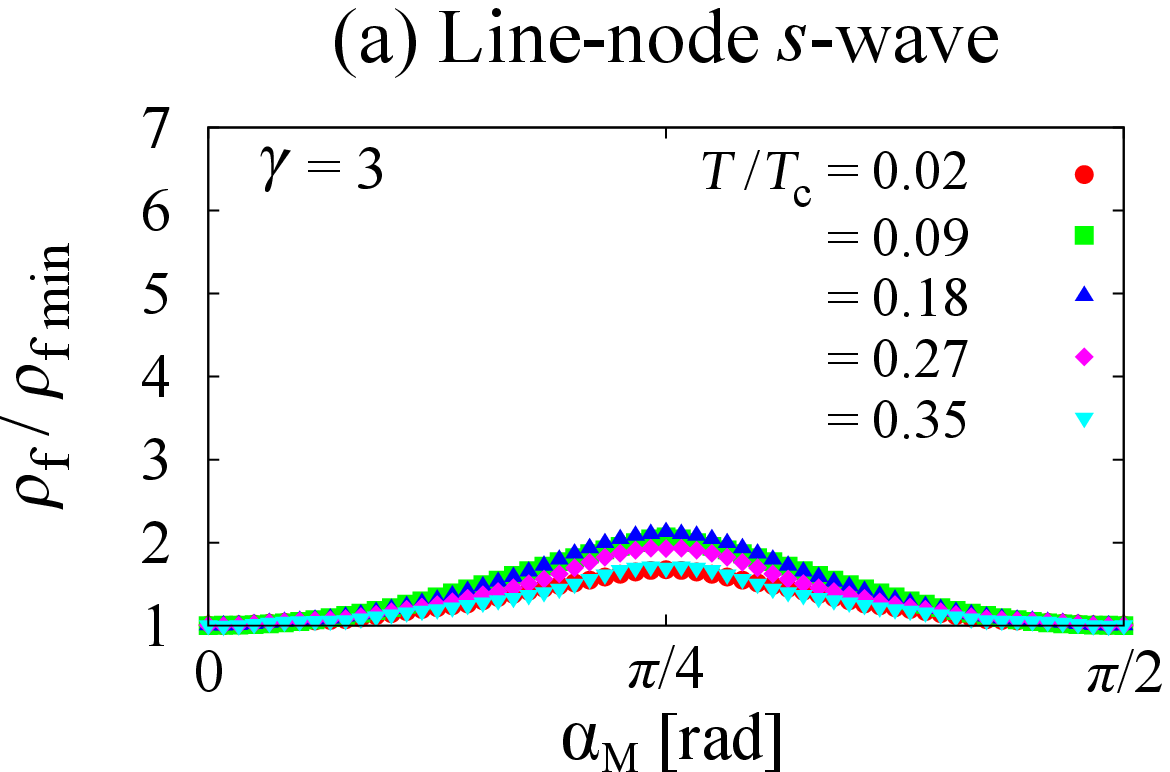}} \\
      \resizebox{70mm}{!}{\includegraphics{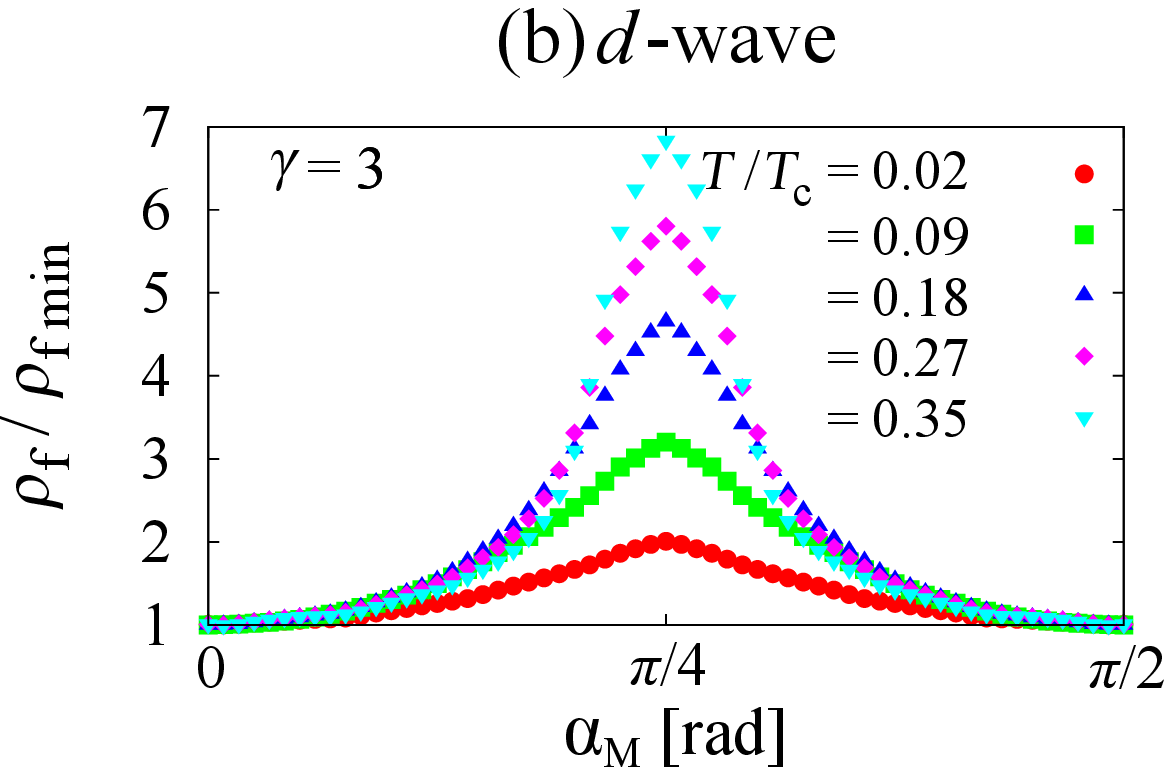}} 
    \end{tabular}
\caption{
\label{fig:fig2}
(Color online)
Field-angle ($\alpha_{\rm M}$) dependence of the flux-flow resistivity $\rho_{\rm f}$
in the case of 
(a) line-node $s$-wave and (b) $d$-wave pair
for a spheroidal FS ($\gamma=3$). 
Each curve indicates different temperature.
The vertical axis is normalized by the minimum value of $\rho_{\rm f~min}$ for each plot.
}
  \end{center}
\end{figure}
\begin{figure}[tb]
  \begin{center}
    \begin{tabular}{p{40mm}p{40mm}p{40mm}p{40mm}}
      \resizebox{40mm}{!}{\includegraphics{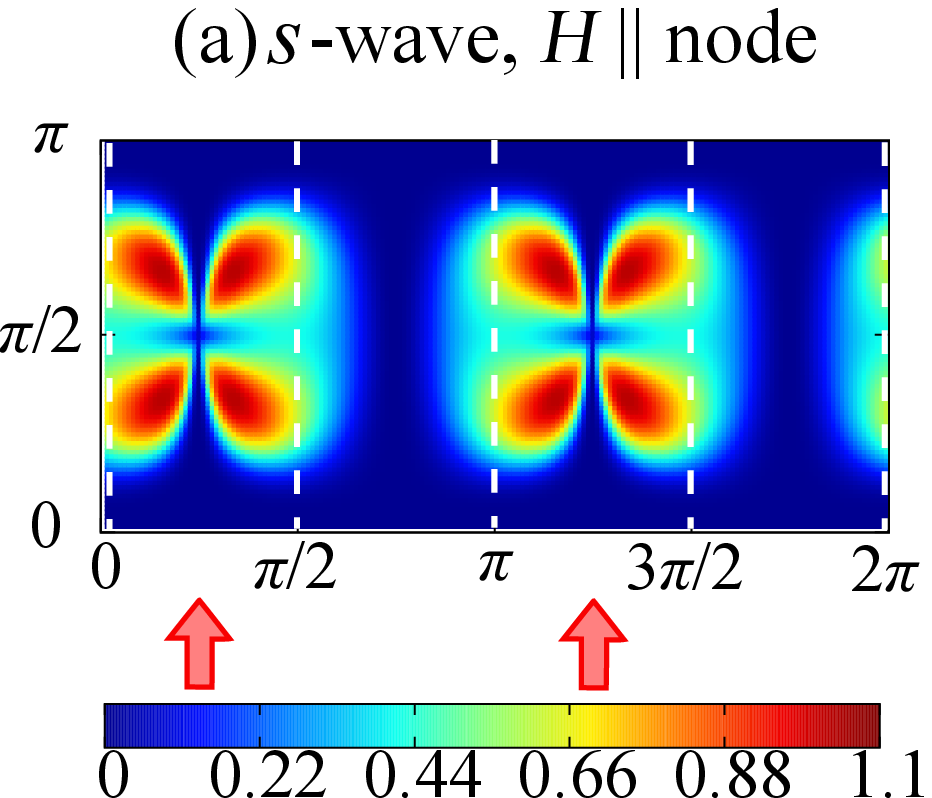}} &
      \resizebox{40mm}{!}{\includegraphics{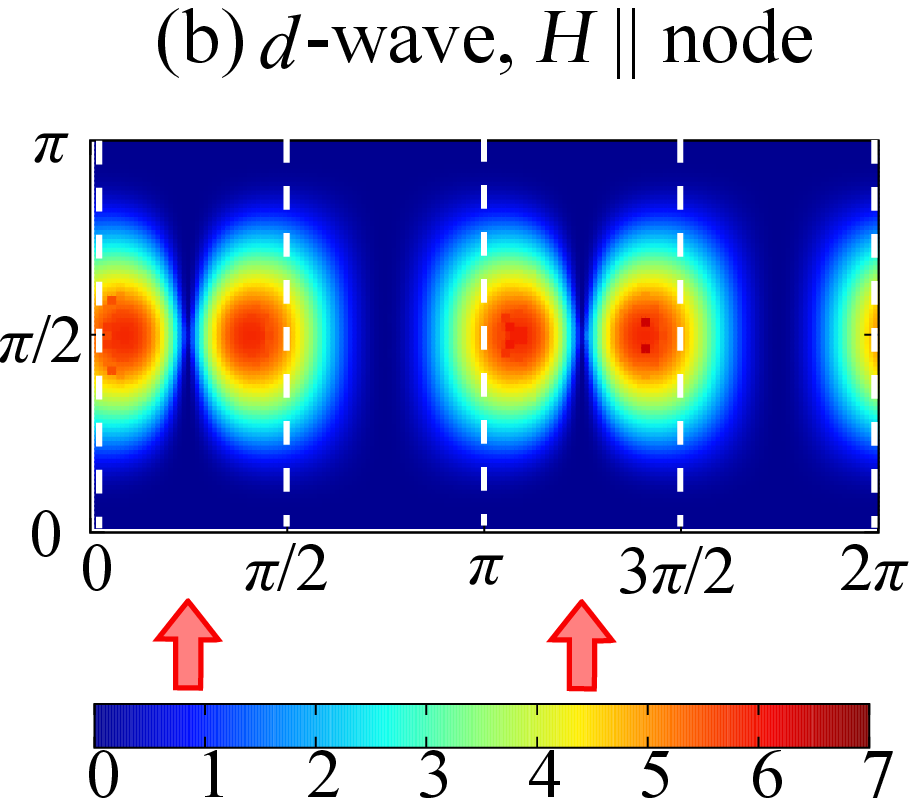}} \\
      \resizebox{40mm}{!}{\includegraphics{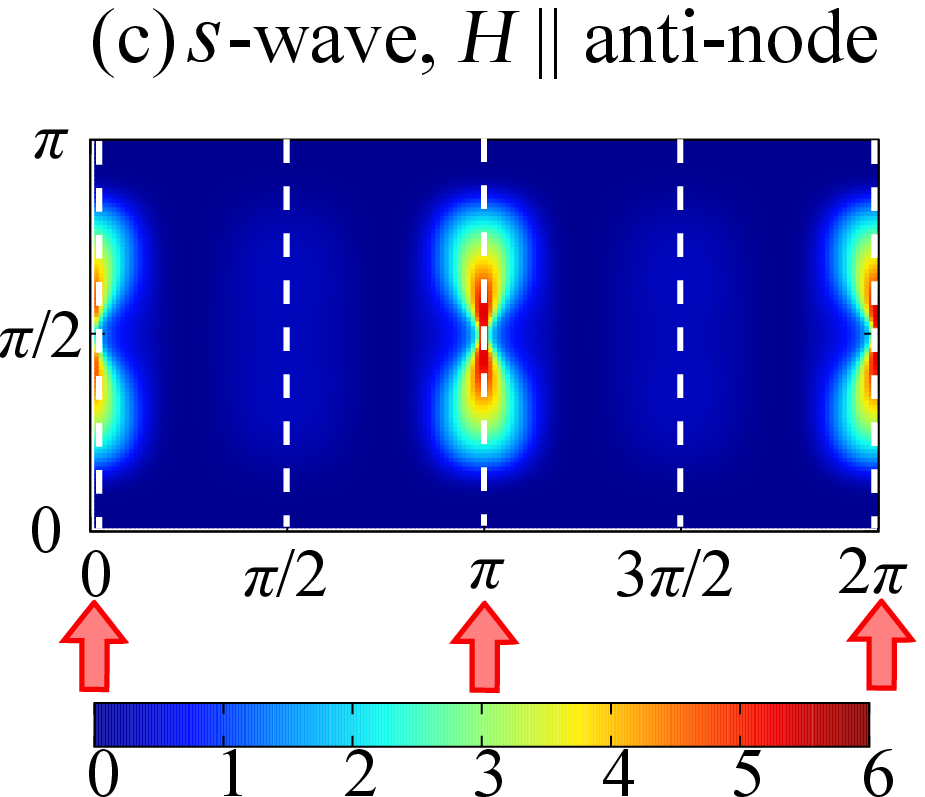}} &
      \resizebox{40mm}{!}{\includegraphics{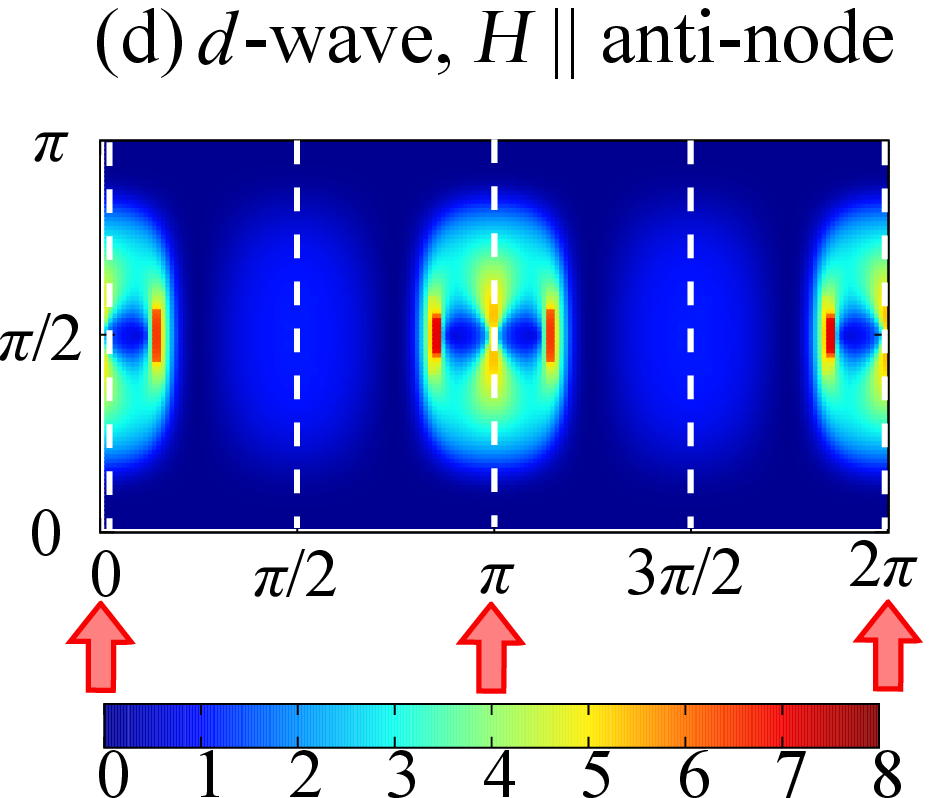}} \\
    \end{tabular}
\caption{
\label{fig:fig3}
(Color online)
$\bm{k}_{\rm F}$ dependence of $\varGamma$ in the case of 
(a) the line-node $s$-wave pair when $\bm{H}~||$ the gap-node direction, 
(b) $d$-wave pair when $\bm{H}~||$ the gap-node direction,
(c) the line-node $s$-wave pair when $\bm{H}~||$ the anti-node direction and 
(d) $d$-wave pair when $\bm{H}~||$ the anti-node direction
on a spheroidal FS ($\gamma=3$). 
The quasiparticle energy $\varepsilon$ is set to $0.2\varDelta_0$ in each plot. 
The vertical and horizontal axes denote the polar angle $\theta_k$ 
and the azimuthal angle $\phi_k$ respectively.
The dotted lines indicate the anit-node directions.
The field directions are indicated by arrows.
}
  \end{center}
\end{figure}

\section{results}
\subsection{Field-angular dependence of flux-flow resistivity}
We show numerical results
for a uniaxially anisotropic FS
with $\gamma=3$.
In Fig.~\ref{fig:fig2}, 
we show the field-angle dependence
of the flux-flow resistivity $\rho_{\rm f}(\alpha_{\rm M})$
for the two pair potential models. 
As shown in Fig.~\ref{fig:fig2}(a),
in the case of the line-node $s$-wave pair,
a broad maximum appears
when $\bm{H}$ is applied
parallel to the gap-node direction 
($\alpha_{\rm M}=\pi/4$).
Note that
the field-angle dependence of the QP scattering rate
$\varGamma(\alpha_{\rm M})$
has its minimum
when $\bm{H}$ is parallel to the node direction.\cite{higashi2011,higashi-iss2011}
The oscillation amplitude of $\rho_{\rm f}(\alpha_{\rm M})$
remains small compared with the $d$-wave case
when the temperature $T$ is increased.
$\rho_{\rm f}(\alpha_{\rm M})$
has little dependence on $T$
in the case of the line-node $s$-wave pair.

On the other hand,
in the $d$-wave case [Fig.~\ref{fig:fig2}(b)], 
a sharp maximum appears
when $\bm{H}$ is applied to the gap-node direction.
The oscillation amplitude
grows with increasing $T$
in contrast to the line-node $s$-wave pair.
This behavior indicates that
the peak of $\rho_{\rm f}(\alpha_{\rm M})$
has a strong temperature dependence
in the $d$-wave case.

The field-angle dependence of $\rho_{\rm f}$
is quite contrasting
between the line-node $s$-wave pair and the $d$-wave one.
One would question what the reason for this prominent difference is.
We consider that
this difference
comes from whether there is a sign change in the pair potential or not.

\subsection{Quasiparticle scattering on the Fermi surface}
First,
we list the characteristics of the $\bm{k}_{\rm F}$ dependence of the QP scattering rate $\varGamma(\bm{k}_{\rm F})/\varGamma_{\rm n}$.
$\varGamma(\bm{k}_{\rm F})/\varGamma_{\rm n}$ is obtained by integrating Eq.~(\ref{k-dep-qp_scattering_rate})
with respect to $\bm{k}^\prime_{\rm F}$.
Next, we explain the behavior of $\varGamma(\alpha_{\rm M})$. 

To clarify
why $\rho_{\rm f}(\alpha_{\rm M})$ behaves contrastingly 
between the two pair potential models, 
we investigate the $\bm{k}_{\rm F}$ dependence of $\varGamma$ first.
$\varGamma(\bm{k}_{\rm F})/\varGamma_{\rm n}$ indicates
which QPs are easy to be scattered on the FS.
We find the following characteristics of the QP scattering:
(i)
The QPs in the vicinity of the anti-node direction
predominantly contribute to $\varGamma$,
as seen in Figs.~\ref{fig:fig3}(a)--3(d).
$\varGamma(\bm{k}_{\rm F})/\varGamma_{\rm n}$
has a higher value around anti-node directions
(see around the dotted lines).
We describe the physical picture for this characteristics as follows.
The QPs flowing in the direction of the gap nodes
feel the small amplitude of the pair potential [i.e., small $\varDelta_0|d(\bm{k}_{\rm F})|$] even in the bulk.
Then, $\xi_{\rm eff}(\bm{k}_{\rm F})=|\bm{v}_{\rm F \perp}|/[\varDelta_0|d(\bm{k}_{\rm F})|]$ becomes large.
Thus a vortex core spreads out effectively because of the large effective coherence length $\xi_{\rm eff}$,
and the QP wave function extends outside a vortex core.
The wave function is damped exponentially by a factor $\exp[-u(s_0,\bm{k}_{\rm F})]$
and $\varGamma$ for the QPs in the node directions becomes small [see Eq.~(\ref{F})].
On the other hand,
the QPs flowing in the direction of the anti-nodes
feel the full amplitude of the pair potential ($\varDelta_0$).
Then, 
a vortex core gets small effectively because $\xi_{\rm eff}$ becomes small.
Hence, the QP wave function is strongly localized inside a vortex core and scattered inside it,
giving a large contribution to $\varGamma$.
(ii)
There is the tendency that the QPs in the direction of $\bm{H}$
are easy to be scattered. 
As seen in Figs.~\ref{fig:fig3}(a) and \ref{fig:fig3}(b),
this property of the QP scattering is confirmed
by the fact that
the weight of $\varGamma$ shifts
a bit toward the field direction.
The  tendency is obvious in Figs.~\ref{fig:fig3}(c) and \ref{fig:fig3}(d).
The reason why the QPs have above tendency is because
$|\bm{v}_{{\rm F}\perp}(\bm{k}_{\rm F})|$
of the QPs in the field direction is small
and the contribution to $\varGamma$ becomes large [see Eq.~(\ref{F}) and Fig.~\ref{fig:fig6}].
(iii)
The other contribution to $\varGamma$ is
expressed by the coherence factor 
$C(\bm{k}_{\rm F},\bm{k}^\prime_{\rm F})$,\cite{Nagai2010}
which reflects the sign of the pair potential in Eq.\ (\ref{coherence-factor}).
This characteristic is discussed in detail in Ref.~\onlinecite{Nagai2010}.
Here we summarize their results (i.e., the dependence of $\varGamma$ on the QP scattering types) in Table~\ref{table:1}.
\begin{center}
\begin{table}[t]
\caption{The QP scattering types.}
\begin{tabular}{ccc}
\hline
\hline
& Forward scattering & Backward scattering\\
& $\Theta(\bm{k}_{\rm F},\bm{k}^\prime_{\rm F})=0$ & $\Theta(\bm{k}_{\rm F},\bm{k}^\prime_{\rm F})=\pi$ \\
\hline
sign-conserved & suppressed &  small\\
sign-reversed & enhanced &  suppressed\\
\hline
\hline
\end{tabular}
\label{table:1}
\end{table}
\end{center}

\section{field-angular dependence of the quasiparticle scattering rate}
First of all, 
we consider $\varGamma$,
which is a part of the contribution to $\rho_{\rm f}$ as shown in Eq.\eqref{flux-flow_resistivity}.
In the case of the line-node $s$-wave pair,
taking into account the characteristics of the QP scattering (i) -- (iii),
we can explain qualitatively the behavior of $\varGamma(\alpha_{\rm M})$.
In the line-node $s$-wave pair,
$\varGamma(\alpha_{\rm M})$
has its minimum when $\bm{H}$ is applied to the gap-node direction \cite{higashi2011,higashi-iss2011}.
In this case,
according to the factor (iii),
the effect of the coherence factor on $\varGamma(\bm{k}_{\rm F})/\varGamma_{\rm n}$ is small.
In addition,
in the line-node $s$-wave pair,
we confirmed that the coherence factor has no field-angle dependence.
Hence,
we can neglect the factor (iii) in the line-node $s$-wave pair.

When $\bm{H}$ is parallel to the gap-node direction,
the weight of $\varGamma(\bm{k}_{\rm F})/\varGamma_{\rm n}$ shifts
toward the field direction (i.e., the gap-node direction) due to the factor (ii).
Therefore,
the QPs around the gap node
become easier to be scattered.
However, 
according to the factor (i),
the contribution of the QPs in the vicinity of the gap nodes to $\varGamma(\bm{k}_{\rm F})/\varGamma_{\rm n}$
is small.
Hence,
the large contribution to $\varGamma(\bm{k}_{\rm F})/\varGamma_{\rm n}$
due to the factor (ii)
gets small due to the factor (i).

On the other hand, 
when $\bm{H}$ is parallel to the anti-node direction,
the QPs in the direction of $\bm{H}$
(i.e.,~the anti-node direction),
which have small $\bm{v}_{{\rm F}\perp}(\bm{k}_{\rm F})$,
can give a large contribution to $\varGamma(\bm{k}_{\rm F})/\varGamma_{\rm n}$
due to the factor (ii).
In this case,
contrary to the case of $\bm{H}$ parallel to the node direction,
$\varGamma$ remains large due to the factor (i).
%
As a result of the above consideration,
the minimum of $\varGamma(\alpha_{\rm M})$
appears when $\bm{H}$ is parallel to the gap-node direction.
\begin{figure}[t]
  \begin{center}
    \begin{tabular}{p{70mm}}
      \resizebox{70mm}{!}{\includegraphics{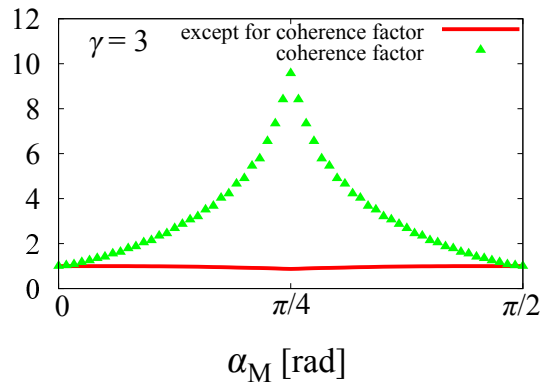}}
    \end{tabular}
\caption{
\label{fig:fig5}
(Color online)
Field-angle ($\alpha_{\rm M}$) dependences of the part of $\varGamma$
containing the coherence factor and the other part of it
for the $d$-wave pair with the spheroidal FS. 
The vertical axis is normalized by the minimum value.
}
  \end{center}
\end{figure}

In the case of the $d$-wave pair,
the backward scattering is suppressed (see Table~\ref{table:1})
since the coherence factor becomes zero
and the scattering point is far from the vortex core.
However, 
the forward scattering with the sign change of pair potential is enhanced (see Table~\ref{table:1}).
So the coherence factor gives the large contribution to $\varGamma(\bm{k}_{\rm F})/\varGamma_{\rm n}$.
In Fig.~\ref{fig:fig5},
we show the field-angle dependences of 
$\langle \langle C(\bm{k}_{\rm F},\bm{k}^\prime_{\rm F})D(\bm{k}_{\rm F},\bm{k}^\prime_{\rm F}) \rangle_{\rm FS^\prime} \rangle_{\rm FS}$,
which is the part of $\varGamma(\varepsilon)/\varGamma_{\rm n}$
containing the coherence factor $C(\bm{k}_{\rm F},\bm{k}^\prime_{\rm F})$.
We also calculate the field-angle dependences of $\langle \langle F(\varepsilon,\bm{k}_{\rm F},\bm{k}^\prime_{\rm F})\rangle_{\rm FS^\prime} \rangle_{\rm FS}$,
which does not contain the coherence factor.
When $\bm{H}$ is parallel to the node direction,
$\langle \langle C(\bm{k}_{\rm F},\bm{k}^\prime_{\rm F})D(\bm{k}_{\rm F},\bm{k}^\prime_{\rm F})\rangle_{\rm FS^\prime} \rangle_{\rm FS}$
shows a sharp maximum.
On the other hand,
$\langle \langle F(\varepsilon,\bm{k}_{\rm F},\bm{k}^\prime_{\rm F})\rangle_{\rm FS^\prime} \rangle_{\rm FS}$
shows little field-angle dependence.
This sharp maximum reproduces the behavior of $\varGamma(\alpha_{\rm M})$ in the $d$-wave case.\cite{higashi2011,higashi-iss2011}

Let us explain the physical picture of the QP scattering rate in the $d$-wave case.
First of all, we should note that the intensity of the forward scattering is the important factor of the QP scattering around a vortex,
since the forward scatterings occur when the scattering point is near the vortex center (i.e., the QP scattering occurs inside a vortex core), as shown in Fig.~\ref{fig:fig1}.
Thus, we consider the field-angle dependence of the intensity of the forward scattering.
We note that the intensity of the forward scattering becomes larger upon decreasing the effective coherence length $\xi_{\rm eff}(\bm{k}_{\rm F})$.
The effective coherence length $\xi_{\rm eff}(\bm{k}_{\rm F})$ defined by Eq.~\eqref{xi_eff} is proportional to the projected Fermi velocity $\bm{v}_{\rm F \perp}(\bm{k}_{\rm F})$
and is inversely proportional to the amplitude of a pair potential $\varDelta_0 |d(\bm{k}_{\rm F})|$.
The minimum effective coherence length is zero when the Fermi velocity $\bm{v}_{\rm F}(\bm{k}_{\rm F})$ is parallel to $\bm{H}$ [i.e., $\bm{v}_{\rm F \perp}(\bm{k}_{\rm F})$ becomes zero]. 
As shown in Fig.~\ref{fig:fig3},
the intensity of $\varGamma(\bm{k}_{\rm F})$ becomes large in the region where $\bm{v}_{\rm F \perp}(\bm{k}_{\rm F})$ becomes small.
This is the reason for the factor (ii).

\begin{figure}[tb]
  \begin{center}
    \begin{tabular}{p{50mm}}
      \resizebox{50mm}{!}{\includegraphics{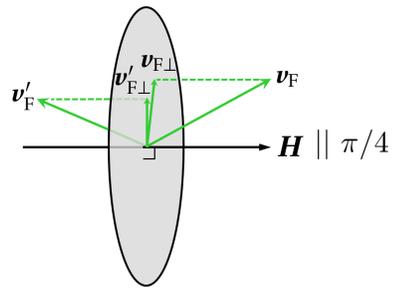}}\\
    \end{tabular}
\caption{
\label{fig:fig6}
Schematic figure of the forward QP scattering,
which is characterized by the angle between the projections of the Fermi velocities onto the plane perpendicular to an in-plane magnetic field $\bm{H}$.
} 
  \end{center}
\end{figure}

The most important factor of the intensity of the forward scattering is the factor (iii). 
As shown in Table~\ref{table:1},
the sign-conserved forward scattering is suppressed even when the effective coherence length becomes small. 
Therefore, the sign-reversed forward scatterings with the small effective coherence length dominantly contribute to the QP scattering rate in the $d$-wave case.
When $\bm{H}$ is parallel to the gap-node direction,
we have confirmed numerically that the forward scattering is realized
by calculating the contribution of anti-nodal QPs to $\varGamma({\varepsilon,\bm{k}^\prime_{\rm F},\alpha_{\rm M}=\pi/4})$.
As shown in Fig.~\ref{fig:fig6},
the forward scattering occurs through the QP scattering in the parallel direction of $\bm{H}$ when $\bm{H}$ is parallel to the gap-node direction.
Moreover, this QP scattering process is sign-reversing, since the quasiparticles with $\bm{v}_{\rm F}$ are scattered across the gap-node perpendicular to $\bm{H}$.
Hence, the sign-reversed forward scattering occurs when $\bm{H}$ is parallel to the gap-node direction even in a single-band superconductor.

On the other hand,
when $\bm{H}$ is parallel to the anti-node direction,
it was revealed through the same analysis that although the sign-reversed scattering occurs,
not only the forward scattering but also the backward scattering occurs.
As a result of the above discussion, 
the quasiparticle scattering rate is enhanced when $\bm{H}$ is parallel to the gap-node direction in the $d$-wave case.

\section{field-angular dependence of flux-flow resistivity}
The other contribution to $\rho_{\rm f}$ is $\omega_0(\bm{k}_{\rm F})$.
When $\varGamma(\varepsilon,\bm{k}_{\rm F})={\rm const.}$,
the field angle dependence of $\omega_0(\bm{k}_{\rm F})$ for a spheroidal FS
is not qualitatively different from that for an isotropic FS.\cite{higashijpcs}
When $\omega_0(\bm{k}_{\rm F})={\rm const.}$,
$\rho_{\rm f}(T) \propto \varGamma(\varepsilon=k_{\rm B}T)$.
In the $d$-wave case,
the dependence of $\omega_0(\bm{k}_{\rm F})$ on the field-angle
makes the behavior of $\varGamma(\alpha_{\rm M})$ stand out.
As a result,
$\rho_{\rm f}(\alpha_{\rm M})$ has a sharp maximum when $\bm{H}$ is parallel to the gap-node direction.
On the other hand,
in the $s$-wave case,
the field angle dependence of $\omega_0(\bm{k}_{\rm F})$ makes
the oscillation amplitude of $\varGamma(\alpha_{\rm M})$ inverted
and $\rho_{\rm f}(\alpha_{\rm M})$ has its broad maximum when $\bm{H}$ is parallel to the gap-node direction.
\section{discussions}
Finally, we comment on the experimental condition for
measuring the flux-flow resistivity under a rotating magnetic field.
Our theory is based on the vortex bound states formed inside a vortex core.
%
Therefore,
an extremely two-dimensional system,
in which a Josephson vortex is formed parallel to the layer,
is beyond our theoretical framework.
However,
we should note that our method can be applied to iron-pnictides, which have a warped cylindrical Fermi surface
such as that found in 11-compounds (FeSe or FeTe) and 122-compounds [BaFe$_2$(As$_{1-x}$P$_x$)$_2$],
since the angular-resolved specific heat and thermal conductivity measurements have successfully detected the gap minima in FeSe$_{0.45}$Te$_{0.55}$ \cite{zeng2010}
and the position of the gap-nodes in BaFe$_2$(As$_{0.67}$P$_{0.33}$)$_2$,\cite{yamashita2011} respectively,
under a rotated magnetic field within the basal plane.
In layered organic compounds $\kappa$-(ET)$_2$Cu(NCS)$_2$,
in-plane field angular dependence of the Josephson-vortex flow resistance
has already been measured by Yasuzuka {\it et al}.,\cite{yasuzuka2010}
but in a three-dimensional system,
measurement of the flux-flow resistivity has not been performed yet.
We calculate $\rho_{\rm f}(\alpha_{\rm M})$ also in the case of the in-plane anisotropic FS.\cite{higashi-iss2012}
In this case,
the behavior of $\rho_{\rm f}(\alpha_{\rm M})$ 
is not qualitatively different from that in the isotropic FS case.
When considering multiband superconductors such as iron-based superconductors,
we need to take into account the contribution from holelike FS to the flux-flow resistivity
in addition to that from electronlike FS.\cite{kopnin1997}
In this study,
we consider the contribution only from electronlike FS.
The multiband effect on the flux-flow resistivity is left for future study.
\section{conclusion}
In conclusion,
we theoretically studied the in-plane magnetic field-angle dependence of the flux-flow resistivity
for a uniaxially anisotropic FS.
We showed that the measurement of the flux-flow resistivity
changing the field direction within the $a$-$b$ plane
can detect both the position of the gap nodes
and the sign change of the pairing potential.
One can estimate the flux-flow resistivity by means of microwaves. 
Instead of fabricating a junction,
one can obtain the information on the phase of the pair potential by measuring the microwave surface impedance under a rotating magnetic field.




\section*{Acknowledgment}
The authors thank T. Okada and S. Yasuzuka for helpful discussions.


\end{document}